\documentclass[a4paper,11pt]{article}
\usepackage{jinstpub}
\usepackage[T1]{fontenc}
\usepackage[utf8]{inputenc}
\usepackage{microtype}
\usepackage{graphicx,booktabs,orcidlink}
\usepackage{tikz}
\usepackage{xcolor}
\usetikzlibrary{arrows.meta,positioning,calc}
\usepackage[section]{placeins}

\title{Graphene Transfer for Vertical Amorphous-Selenium Photodetectors: Fabrication, Optical Transmission and Operation at 77 K}

\author[a,1]{I.~Tzoka\orcidlink{0000-0001-7811-5068}\note{Corresponding author.},}
\author[c]{M. Boukhicha,\orcidlink{orcid.org/0000-0002-8866-5714}}
\author[a]{V.~A.~Chirayath\,\orcidlink{0000-0003-2450-2753},}
\author[a]{B.~Brown,
\orcidlink{0009-0005-0214-3739},}
\author[a]{M. Rooks,\orcidlink{0000-0002-1629-6825}} 
\author[d,2]{M.~Febbraro,\orcidlink{0000-0002-0347-2260}\note{Current address: Air Force Institute of Technology, WPAFB, OH, USA.},}
\author[g]{S. Park,\orcidlink{0000-0002-2269-7705 }}
\author[b]{S. Abbaszadeh,\orcidlink{0000-0001-6345-2408}}
\author[b]{K. Hellier,\orcidlink{0000-0002-6328-1167}}
\author[a]{J. Asaadi,\orcidlink{0000-0001-6915-5279}}
\author[f]{and A.~K.~Rumaiz\,\orcidlink{0000-0002-9907-299X}}
\affiliation[a]{Department of Physics, University of Texas at Arlington}
\affiliation[b]{University of California Santa Cruz, Electrical and Computer Engineering, Santa Cruz, CA 95064, USA}
\affiliation[c]{Instrumentation Department, Brookhaven National Laboratory, Upton, NY, 11973, USA}
\affiliation[d]{Oak Ridge National Laboratory, Physics Division, Oak Ridge, TN 37831, USA}
\affiliation[f]{NSLS-II, Brookhaven National Laboratory, Upton, NY, 11973, USA}
\affiliation[g]{Center for Functional Nanomaterials, Brookhaven National Laboratory, Upton, New York 11973, USA}

\emailAdd{iakovos.tzoka@uta.edu}
\hypersetup{
    pdftitle={Graphene Transfer for Vertical Amorphous-Selenium Photodetectors},
    pdfauthor={I. Tzoka et al.},
    pdfsubject={Graphene transfer, transmission and vertical a-Se device operation},
    pdfkeywords={amorphous selenium, graphene, photodetector, cryogenic, transmission}
}

\abstract{
Graphene can serve as a VUV transparent electrode in vertical photodetector devices. Using an optically transparent graphene as a top-electrode of amorphous selenium (a-Se) photoconductor requires transfer process compatible with the thermal and chemical sensitivities of a-Se thin films. The baking and polymer-removal steps of conventional commercial Trivial Transfer graphene procedures are undesirable because heating can promote selenium crystallization and prolonged solvent exposure can cause delamination. We implement two alternatives: PMMA-free transfer of nominal 50-layer graphene  and transfer of nominal 10-layer graphene using an approximately 80~nm PMMA support removed in acetone at $40\,^{\circ}\mathrm{C}$. Raman G and 2D bands identify graphene after transfer. 
Separate graphene/MgF$_2$ specimens are characterized by dark-corrected transmission measurements over 120--325~nm and compared with earlier single-layer and nominal 6--8-layer Trivial Transfer spectra. The completed graphene/a-Se/Cr devices produce bias-dependent voltage transients under pulsed 415~nm illumination at 293~K and 77~K. These results establish an a-Se-compatible fabrication route for graphene-electrode devices and demonstrate vertical device operation at both room and cryogenic temperatures. The completed devices provide a platform for detection-efficiency studies, and direct VUV photoresponse measurements now underway.
}

\keywords{amorphous photoconductors, graphene, photodetector, cryogenic detectors, Noble liquid detectors}

\begin{document}
\maketitle
\flushbottom

\section{Introduction}
\label{sec:intro}

Liquid argon (LAr) and liquid xenon (LXe) are widely used active media in detectors for neutrino physics and rare-event searches, including searches for dark matter and neutrinoless double beta decay \cite{PDG2024_NonAcceleratorDetectors,Gonzalez-Diaz:2017gxo}. Both liquids produce ionization electrons and scintillation photons following energy deposition. Time projection chambers (TPCs) exploit these complementary signals to reconstruct particle interactions: an applied electric field transports the ionization charge to a readout plane, while photosensors record the scintillation light~\cite{Bonivento:RevModPhys.96.045001,Aprile:RevModPhys.82.2053}. The dominant scintillation emission lies in the vacuum-ultraviolet (VUV), near 128~nm in LAr and 175~nm in LXe~\cite{Heindl2010Scintillation,Li_2026}.

Efficient photon detection contributes directly to the physics reach of these instruments. Prompt scintillation provides an event-time reference that, together with the charge drift time, allows reconstruction of the interaction position along the drift direction~\cite{Aprile:RevModPhys.82.2053}. Light also supports triggering and energy reconstruction. In LXe, combining the anticorrelated scintillation and ionization signals can improve energy resolution~\cite{Aprile2007Anticorrelation}. In LAr, the different scintillation time profiles of electronic and nuclear recoils provide a basis for pulse-shape discrimination~\cite{Lippincott:PhysRevC.78.035801}. Photon collection therefore helps determine the available timing, calorimetric, and background-discrimination information.

Detecting this light requires optical access and photosensor sensitivity at the scintillation wavelengths, together with reliable operation at cryogenic temperatures. Entrance windows, surface layers, and electrodes must be chosen so that the VUV photons reach the active material. Large detectors additionally require light-collection systems that can cover substantial areas while maintaining manageable readout complexity. These requirements have motivated both wavelength-shifting systems coupled to visible-sensitive sensors and photosensors designed for direct VUV detection~\cite{AlvarezGarrote2024,Alvarez-Garrote2024Measurement}.

In LAr detectors, wavelength shifters such as 1,1,4,4-tetraphenyl butadiene (TPB) absorb the VUV scintillation and re-emit at longer wavelengths that can be detected by visible-sensitive photomultiplier tubes (PMTs) or silicon photomultipliers (SiPMs). Polyethylene naphthalate (PEN) films provide an alternative that can ease large-area installation. Measurements of TPB- and PEN-based reflectors in LAr demonstrate the importance of the coating, substrate, and optical coupling for the resulting light yield~\cite{Araujo2022}. More elaborate collectors, such as the X-ARAPUCA developed for DUNE, combine wavelength conversion, dichroic filtering, and photon trapping in a light guide to collect photons over an area larger than the SiPM area~\cite{AlvarezGarrote2024}. These are established approaches whose performance depends on the complete optical assembly.

Direct VUV detection is also an established option, using VUV-sensitive PMTs and SiPMs~\cite{Aprile:RevModPhys.82.2053,Alvarez-Garrote2024Measurement}. For example, measurements of Hamamatsu VUV4 SiPMs demonstrate sensitivity to 127~nm LAr scintillation at cryogenic temperature~\cite{Alvarez-Garrote2024Measurement}. Recent comparisons of VUV-sensitive SiPMs in the LoLX detector further show that their effective response in LXe depends on incidence angle and surface structure as well as on the intrinsic sensor response~\cite{Li_2026}. These results underline the need to characterize a photosensor under its intended optical and thermal conditions. They also motivate investigation of complementary direct-conversion materials and electrode architectures, including structures that could be compatible with large-area or integrated readout.

Amorphous selenium (a-Se) is a candidate material for direct VUV detection \cite{Rooks_2023}. a-Se is an established photoconductor for direct-conversion X-ray imaging, and its deposition onto electronic readout structures has been demonstrated with monolithically integrated CMOS detectors~\cite{Han_2023}. Of particular relevance to noble-liquid detectors, lateral windowless a-Se devices with interdigitated electrodes have demonstrated visible and VUV photoresponse under cryogenic conditions~\cite{Rooks_2023,Rooks_2026}. The more recent study includes direct detection of narrowband 130~nm illumination~\cite{Rooks_2026}. These measurements establish a basis for investigating a-Se as a scintillation-sensitive material, while leaving the performance of each new device geometry to be measured.

A vertical structure places the photoconductive film between overlapping top and bottom electrodes, allowing bias to be applied uniformly across its thickness. This geometry requires a top contact that provides electrical access while allowing light to reach the selenium. Graphene is a candidate electrode with its use as a VUV transparent contact demonstrated in an AlN-based VUV photovoltaic detector~\cite{Zheng2018}. Applying graphene to a-Se introduces a fabrication challenge. Commercial Trivial Transfer graphene uses a PMMA handling layer, and the manufacturer’s procedure includes a 100 °C bake before polymer removal~\cite{acsmaterial_ttg,ACSTTGInstructions}. Such heating is undesirable for a-Se, for which thermal crystallization has been observed in evaporated films~\cite{Ozenbas1986}. Illumination and heating can also alter the local structure of deposited selenium~\cite{Lu2024}. The transfer procedure must therefore be adapted to the photoconductor, rather than adopted solely on the basis of graphene transparency.

Earlier work by our group explored a vertical a-Se structure with an open metal top contact supplemented by transferred graphene~\cite{Tzoka_et_al}. In this work, we present graphene/MgF$_2$ transmission measurements for commercial single-layer and nominal 6--8-layer Trivial Transfer films together with nominal 10-layer and 50-layer films prepared using adapted transfer routes. To address the processing limitation, we adapt two transfer procedures for direct placement of graphene onto a-Se. The resulting nominal 50-layer (50L) and 10-layer (10L) films are identified by Raman spectroscopy and tested as the top electrodes of vertical devices.

The scope is fabrication feasibility, VUV-UV optical characterization of separate electrode specimens, and experimental demonstration of device functionality. We describe the processing rationale and transfer procedures, present Raman and electrical evidence for transferred graphene, and report optical-transmission measurements for all four graphene thicknesses. Finally, we demonstrate photoresponse under pulsed 415~nm illumination at room temperature and 77~K. Transmission through graphene/Mg$F_2$ windows and visible-light operation of the completed devices are complementary measurements. Direct VUV response of the vertical detectors remains to be established.

\section{Device fabrication and graphene transfer}
\label{sec:fabrication}

A vertical a-Se device places the photoconductive film between electrodes on its opposing faces. In the ideal planar geometry, the applied bias directs the electric field through the film thickness, approximately normal to the substrate within the electrode-overlap region. This arrangement provides a defined separation between the electrodes and differs from the side-by-side contacts of a lateral device. While lateral devices can operate windowless, the vertical geometry benefits from improved an electric field uniformity but requires a VUV transparent, electrically conductive window~\cite{Mirzanezhad2025Spectral}. The layered architecture builds on established a-Se detector designs, Zhao and Rowlands investigated a-Se coupled to an active-matrix readout for direct-conversion X-ray imaging~\cite{zhao1995x}, Scheuermann \textit{et al.} developed a solid-state a-Se optical detector with opposing Cr electrodes and carrier-blocking layers~\cite{Scheuermann2015SolidState} and Mirzanezhad \textit{et al.} studied vertical a-Se and Se--Te photodetectors alongside lateral structures~\cite{Mirzanezhad2025Spectral}.

For optical detection, the electrode in the illumination path must also allow photons to reach the photoconductor. Earlier implementations illustrate this requirement using a semitransparent Cr entrance electrode~\cite{Scheuermann2015SolidState} or illumination through a glass/indium-tin-oxide substrate~\cite{Mirzanezhad2025Spectral}. Transmission through the substrate, contacts, and any intervening layers becomes especially important at short wavelengths. For detection of noble-liquid VUV scintillation, this motivates an electrode that combines electrical continuity with optical access at the wavelengths of interest.

Here, transferred graphene itself forms the top electrode of a graphene/a-Se/Cr stack, with light incident from the graphene side. The fabrication procedures below adapt the established vertical geometry to this contact scheme. Section~\ref{sec:transmission} characterizes separate graphene/MgF$_2$ optical specimens, while Section~\ref{sec:photoresponse} tests the completed vertical devices under 415~nm illumination.

\subsection{Processing constraint and transfer strategy}
\label{sec:constraint}

The thermal sensitivity of the selenium substrate motivates adapting the graphene-transfer procedure. The commercial Trivial Transfer Graphene has an approximately 500~nm PMMA support~\cite{acsmaterial_ttg}. Its instructions specify drying followed by a $100\,^{\circ}\mathrm{C}$ bake for 20~min, then PMMA removal in acetone preheated to $50\,^{\circ}\mathrm{C}$~\cite{ACSTTGInstructions}. These instructions allow adjustment for the target substrate, nevertheless, following that thermal sequence on a-Se is undesirable because preserving the amorphous photoconductor is a fabrication requirement.

Thermal crystallization has been reported for evaporated selenium films annealed between 40 and $90\,^{\circ}\mathrm{C}$~\cite{Ozenbas1986}. Those conditions are not a universal crystallization threshold: composition, film preparation and exposure time differ from the stabilized films used here. They establish why an additional transfer bake cannot be assumed harmless. Our transfer development therefore sought to avoid the commercial $100\,^{\circ}\mathrm{C}$ bake and reduce the polymer-removal burden after graphene placement. Trials with thinner PMMA supports also produced delamination during extended acetone exposure or no detectable graphene Raman features after processing. These observations motivated a PMMA-free route for the 50L film and an approximately 80~nm PMMA support for the 10L film. The resulting procedures demonstrate successful transfer and subsequent device operation, a quantitative comparison of selenium crystallization before and after each procedure has not been performed.

\subsection{Vertical device structure}

The devices were fabricated on $2.5~\mathrm{cm}\times2.5~\mathrm{cm}\times1.1~\mathrm{mm}$ soda lime glass slides from MSE Supplies. A 500~nm Cr bottom electrode was deposited by DC magnetron sputtering at the University of Texas at Arlington Shimadzu Nanofabrication Center. A nominally 10~$\mu$m a-Se film was then deposited by thermal evaporation at Brookhaven National Laboratory. The selenium source had a nominal purity of 99.995\% and was stabilized with 0.2 wt\% As and 10~ppm Cl. Deposition was performed at approximately 90~\AA/s, with a base pressure of approximately $2\times10^{-6}$~Torr, a deposition pressure of approximately $3\times10^{-4}$~Torr, and a substrate temperature of $60\,^{\circ}\mathrm{C}$. After deposition, the substrate was held in the dark at $40\,^{\circ}\mathrm{C}$ for 24~h. Film thickness was measured using a Bruker DektakXT stylus profilometer.

Graphene was transferred onto the selenium to complete the graphene/a-Se/Cr stack. The two devices studied here differ in their nominal graphene layer count, transfer procedure, and electrode area (Table~\ref{tab:devices}). The a-Se films were deposited in the same run. The labels 10L and 50L refer to the nominal layer counts of the starting chemical-vapor-deposited (CVD) graphene films.

\begin{table}[htbp]
    \centering
    \begin{tabular}{lcccc}
        \toprule
        Device & Cr thickness & a-Se thickness & Graphene & Graphene dimensions \\
        \midrule
        Sample~I & 500~nm & 10~$\mu$m & nominal 50L & $13\times13$~mm$^2$ \\
        Sample~II & 500~nm & 10~$\mu$m & nominal 10L & $6\times10$~mm$^2$ \\
        \bottomrule
    \end{tabular}
    \caption{Layer structure and graphene dimensions of the two devices.}
    \label{tab:devices}
\end{table}

\subsection{Graphene-transfer procedures}

Sample~I used nominal 50L graphene grown on Ni and a modified bifacial float-transfer procedure based on Andrade \textit{et al.}~\cite{https://doi.org/10.1002/adfm.202005103}. The starting structure comprised a PET/EVA support and graphene on both sides of the Ni foil. The foil was submerged in an aluminum etchant solution to release the graphene film. The solution was gradually diluted to bring the detached film to the liquid surface, followed by successive exchanges with deionized water. The graphene was then transferred onto the a-Se surface. This procedure did not require a PMMA support layer.

Sample~II used nominal 10L graphene grown on Cu. An approximately 80~nm PMMA support film was spin-coated onto the graphene with its thickness measured using a Filmetrics F-40 reflectometer. The supported film was floated on a 1~M ammonium persulfate solution until the Cu was etched, rinsed with deionized water, and transferred onto a-Se. The PMMA was removed by immersion in acetone at $40\,^{\circ}\mathrm{C}$ for 30~min, followed by drying under ambient conditions. The thin support was selected empirically to permit handling during Cu removal while limiting the subsequent solvent treatment. Neither transfer sequence included the $100\,^{\circ}\mathrm{C}$ post-placement bake specified in the commercial procedure. The selenium-deposition and dark-conditioning temperatures stated above describe earlier film preparation and are distinct from the graphene-transfer sequence.

\begin{figure}[htbp]
    \centering
\begin{tikzpicture}[
    x=1cm,y=1cm,font=\small,
    flowbox/.style={draw=black!55,rounded corners=2pt,fill=black!2,
        text width=5.5cm,minimum height=1.05cm,align=center,inner sep=5pt},
    flow/.style={-{Stealth[length=2mm]},thick,draw=black!65}]
    \node[font=\bfseries] at (-3.35,0.25) {Sample I: nominal 50L};
    \node[font=\bfseries] at (3.35,0.25) {Sample II: nominal 10L};
    \node[flowbox] (i1) at (-3.35,-0.8) {Graphene on Ni\\PET/EVA support stack; no PMMA};
    \node[flowbox] (ii1) at (3.35,-0.8) {Graphene on Cu\\Spin-coat $\sim$80 nm PMMA};
    \node[flowbox] (i2) at (-3.35,-2.25) {Release film in etchant\\Dilute and exchange with DI water};
    \node[flowbox] (ii2) at (3.35,-2.25) {Etch Cu in 1 M\\ammonium persulfate; DI-water rinse};
    \node[flowbox] (i3) at (-3.35,-3.7) {Transfer graphene onto a-Se\\No PMMA-removal step};
    \node[flowbox] (ii3) at (3.35,-3.7) {Transfer onto a-Se; remove PMMA\\Acetone, $40\,^{\circ}\mathrm{C}$, 30 min; air dry};
    \draw[flow] (i1)--(i2); \draw[flow] (i2)--(i3);
    \draw[flow] (ii1)--(ii2); \draw[flow] (ii2)--(ii3);
    \node[font=\small\itshape,align=center] at (0,-4.8)
        {Both routes omit the commercial $100\,^{\circ}\mathrm{C}$ transfer bake.};
    \node[font=\bfseries] at (0,-5.65) {Completed vertical device and measurement connections};
    \fill[black!7] (-2.5,-7.9) rectangle (2.5,-7.45);
    \draw[black!55] (-2.5,-7.9) rectangle (2.5,-7.45);
    \fill[black!45] (-2.5,-7.45) rectangle (2.5,-7.23);
    \draw[black!70] (-2.5,-7.45) rectangle (2.5,-7.23);
    \fill[orange!18] (-2.5,-7.23) rectangle (2.5,-6.5);
    \draw[black!55] (-2.5,-7.23) rectangle (2.5,-6.5);
    \draw[line width=2pt,blue!65!black] (-2,-6.48)--(2,-6.48);
    \node at (0,-7.68) {Glass substrate};
    \node at (0,-6.88) {a-Se: $\sim$10 $\mu$m};
    \node[anchor=west] at (2.65,-7.35) {Cr: 500 nm};
    \node[anchor=east] at (-2.65,-6.48) {Graphene};
    \draw[flow,blue!65!black] (-0.7,-5.95)--(-0.7,-6.4);
    \node[anchor=east,font=\scriptsize] at (-0.85,-6.05) {415 nm light};
    \draw[flow] (1.75,-6.48)--(3.5,-6.48);
    \node[anchor=west,font=\scriptsize,align=left] at (3.5,-6.48)
        {AC-coupled voltage readout\\$1~\mathrm{M}\Omega$ input};
    \draw[flow] (-4.3,-7.34)--(-2.55,-7.34);
    \node[anchor=east,font=\scriptsize] at (-4.3,-7.34) {$+V_{\mathrm{bias}}$};
\end{tikzpicture}
    \caption{Transfer procedures and completed vertical device geometry. Sample~I uses a PMMA-free float transfer of nominal 50L graphene from Ni; Sample~II uses nominal 10L graphene from Cu with an approximately 80~nm PMMA support. Both routes omit the commercial $100\,^{\circ}\mathrm{C}$ post-placement bake. The device schematic indicates the bias and readout configuration used for photoresponse measurements; dimensions are not to scale.}
    \label{fig:process}
\end{figure}
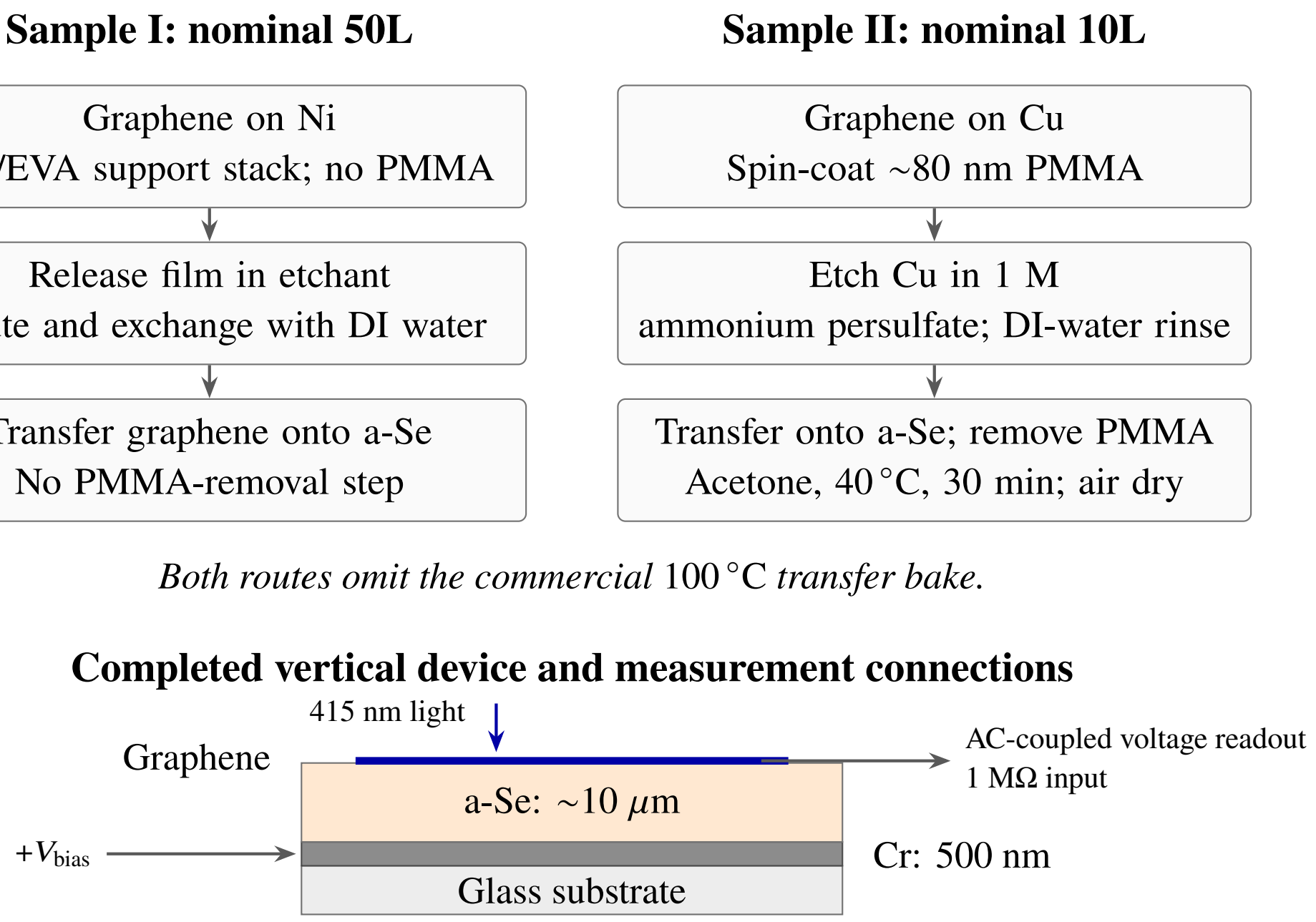

Figure~\ref{fig:process} summarizes the two transfer procedures. Silver-epoxy contacts were applied to the transferred graphene and allowed to cure for 24~h before electrical testing. These contacts provided a surface for the tungsten probes without probing directly through the graphene film.

\section{Evidence for successful graphene transfer}
\label{sec:transfer_evidence}

\subsection{Raman spectroscopy}

Raman spectra were acquired at the Center for Functional Nanomaterials at Brookhaven National Laboratory using a WITec alpha300 RA Raman--AFM microscope. Measurements used a 50X objective, 532~nm excitation, a nominal laser power of 2~mW, and 10 accumulations of 5~s each.

\begin{figure}[htbp]
    \centering
    \begin{minipage}[b]{0.47\textwidth}
        \centering
        \includegraphics[width=\linewidth]{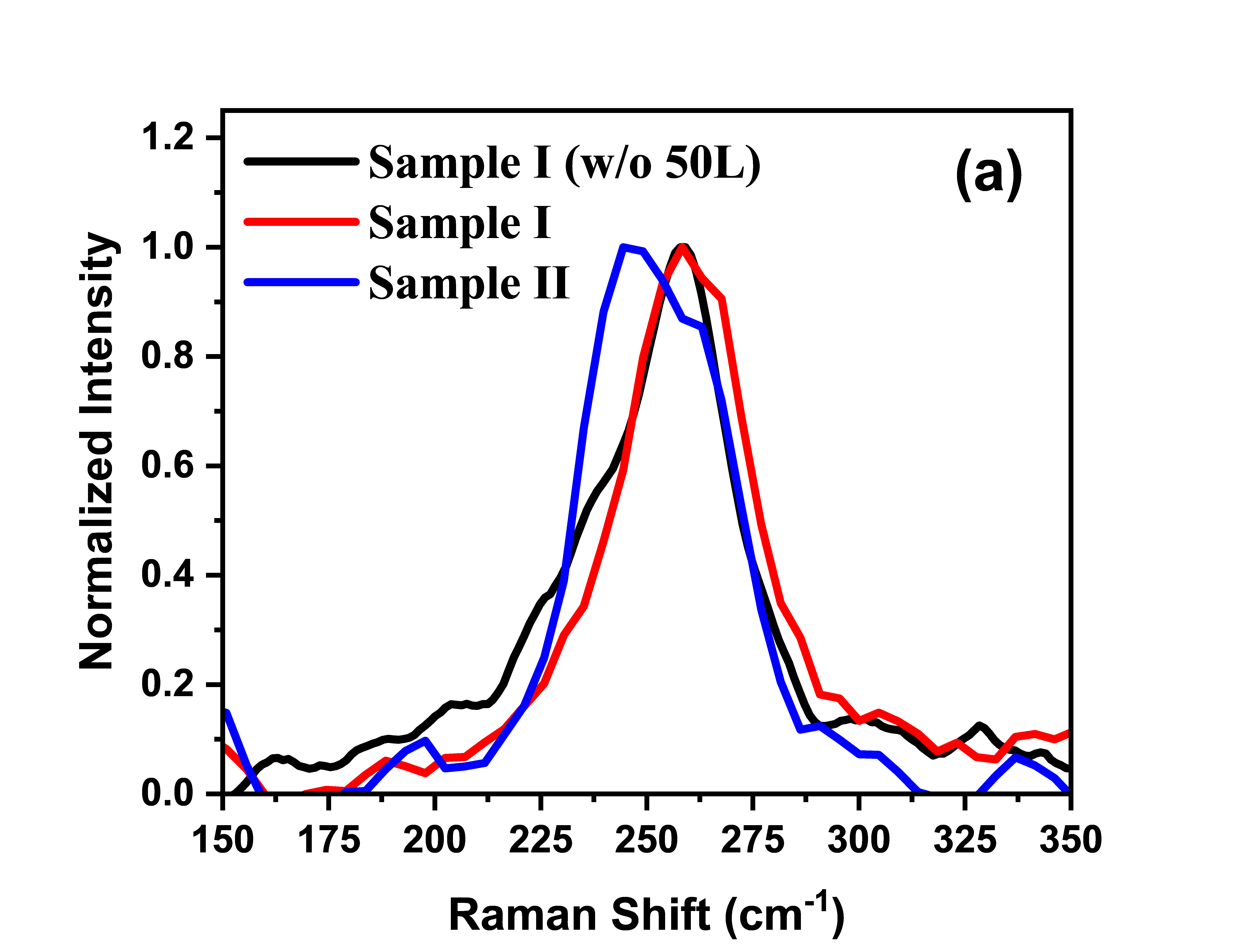}
    \end{minipage}
    \hfill
    \begin{minipage}[b]{0.47\textwidth}
        \centering
        \includegraphics[width=\linewidth]{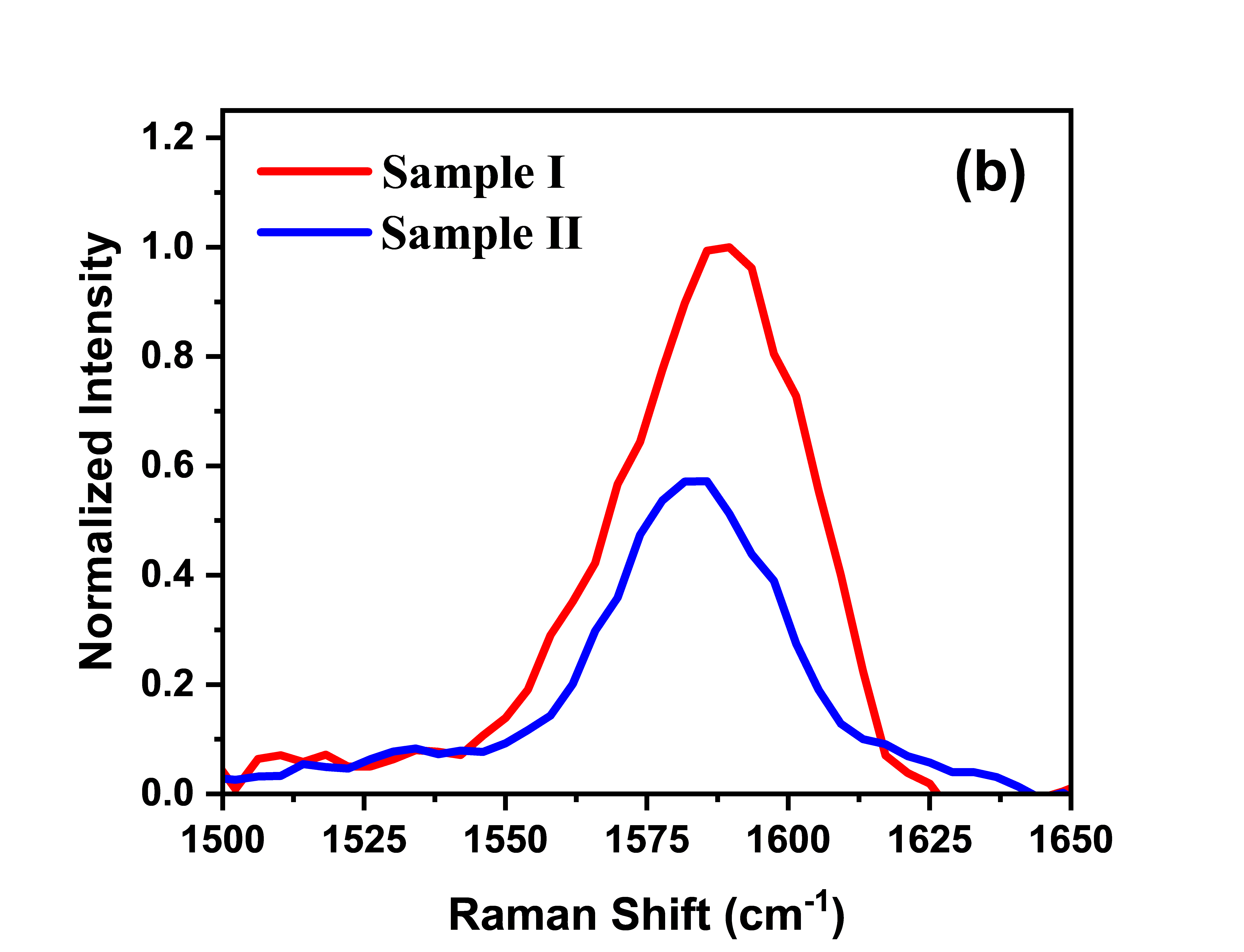}
    \end{minipage}

    \medskip
    \begin{minipage}[b]{0.47\textwidth}
        \centering
        \includegraphics[width=\linewidth]{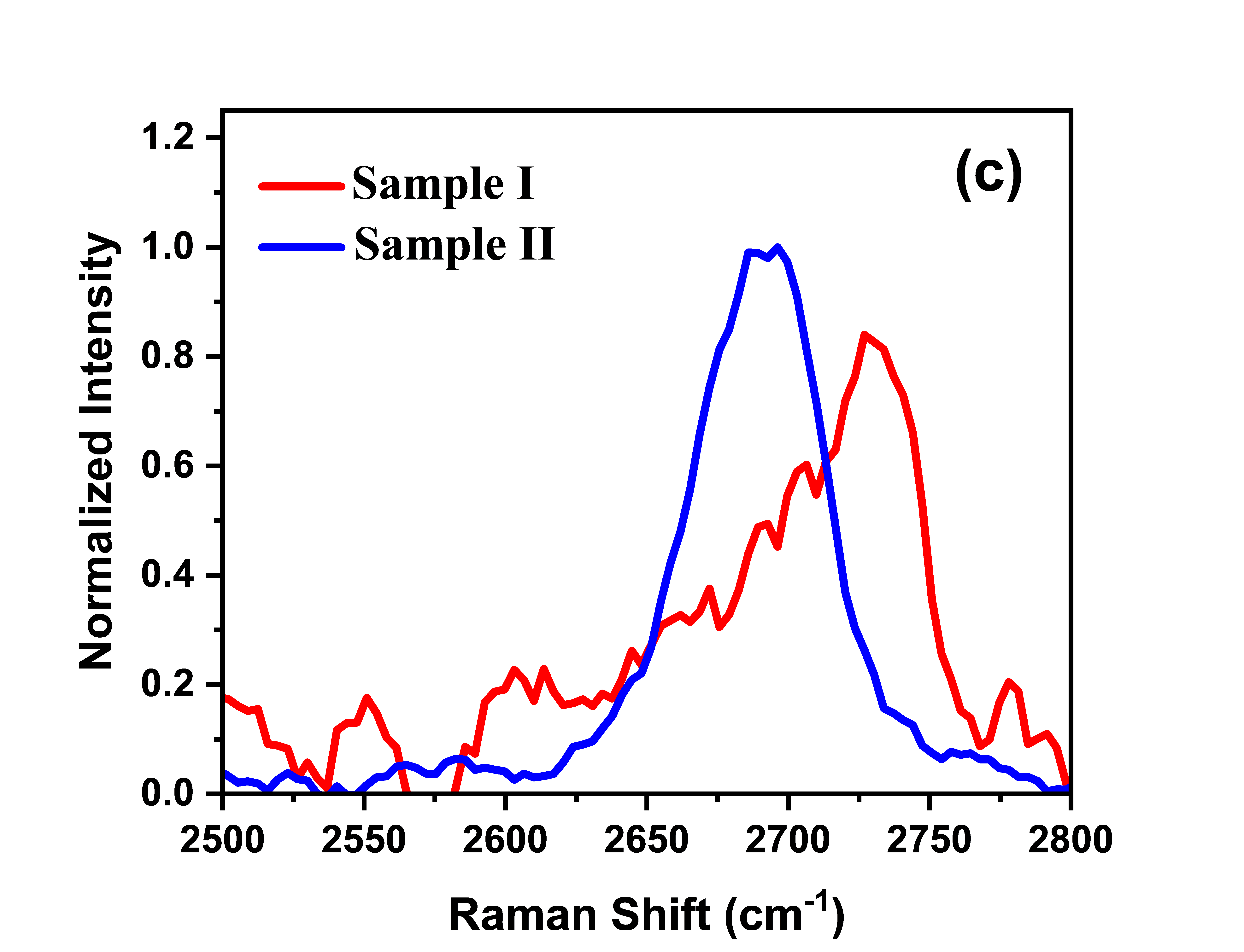}
    \end{minipage}
    \caption{Raman spectra after graphene transfer. (a) Selenium region for Samples~I and II and an uncovered region of Sample~I processed alongside the graphene-covered region. (b) Graphene G-band and (c) 2D-band regions for both samples. The G and 2D features establish graphene presence after transfer. The selenium spectra show local differences but do not distinguish transfer-induced changes from changes under Raman illumination. The 50L and 10L labels denote nominal starting-film layer counts.}
    \label{fig:raman}
\end{figure}

Both devices exhibit graphene G and 2D bands following transfer, near 1580 and 2700~cm$^{-1}$, respectively (Fig.~\ref{fig:raman}). These characteristic features identify graphene on the processed a-Se surface~\cite{Ferrari2013}. Sample~II has a comparatively narrow 2D feature, while Sample~I has a broader, asymmetric feature. Because Raman line shapes also depend on stacking, strain, doping, and disorder, these spectra are used to establish graphene presence; they do not independently determine the nominal 10L and 50L layer counts.

The selenium Raman region is shown in Fig.~\ref{fig:raman}(a), including an uncovered region of Sample~I that underwent the same transfer processing. Broad spectral weight near 250~cm$^{-1}$ is consistent with amorphous/disordered selenium, while contributions near 233--237~cm$^{-1}$ can indicate trigonal selenium~\cite{Poborchii1998,Lu2024}. The line shapes differ between the measured regions. Above-bandgap Raman excitation can itself modify vapor-deposited a-Se~\cite{Lu2024}, so the present spectra do not separate changes caused by transfer from changes during spectroscopy. They support the presence of an amorphous/disordered selenium contribution after processing, while leaving local structural modification unresolved. Accordingly, successful graphene transfer is not taken to establish that the underlying selenium is structurally unchanged.

\subsection{Electrical continuity}

Two-terminal resistance measurements with multimeter probes placed across the long edges of the transferred films gave approximately 2.4--2.5~k$\Omega$ for both graphene electrodes. These measurements establish electrical continuity across the films. They do not independently measure sheet resistance or graphene/a-Se contact resistance, nor do they establish complete removal of PMMA residue from Sample~II. Together with the Raman identification, electrical continuity provides evidence that the transfer procedures yield graphene films suitable for testing as top electrodes.

\section{Optical transmission of graphene electrode specimens}
\label{sec:transmission}

\subsection{Measurement setup and comparison samples}

We measured optical transmission for four sets of graphene-coated MgF$_2$ windows: commercial single-layer and nominal 6--8-layer Trivial Transfer films, and nominal 10L and 50L films prepared using the transfer routes developed for a-Se. These optical specimens were measured separately from the completed graphene/a-Se/Cr devices.

The commercial single-layer and nominal 6--8-layer CVD graphene films were obtained from ACS Material~\cite{acsmaterial_ttg}. Transfers onto MgF$_2$ were performed in a Class-100 laminar flow hood. The PMMA-supported graphene was floated on anhydrous ethanol during transfer to limit water exposure of the MgF$_2$ surface. After placement on the window, the PMMA support was removed using acetone. A bare MgF$_2$ control window underwent the same solvent-handling procedure without graphene transfer. The windows were stored under low-humidity conditions before preparation and mounted in the vacuum instrument after transfer. The nominal 10L and 50L films were transferred onto MgF$_2$ using the same thin-PMMA-assisted and PMMA-free routes, respectively, as those used on selenium in Section~\ref{sec:fabrication}.The bare MgF2 used as the control window for 10L and 50L transmission measurements was not treated under the same solvent-handling procedure and was mounted to the monochromator setup in the as purchased form.

Transmission measurements were performed using a Resonance Ltd. VM200 monochromator and a deuterium lamp with an MgF2 output window. The single-layer and nominal 6-8 layer samples were measured at Oak Ridge National Laboratory (ORNL). Transmitted light was detected with an OSI Optoelectronics XUV-100 VUV photodiode connected to a transimpedance amplifier, and the amplifier output was recorded using a Keithley 6510 DAQ/multimeter. A VUV bandpass
filter was used to remove higher order terms from the diffraction
grating. The chamber pressure was maintained below $10^{-6}$ Torr using a turbomolecular pump backed by an oil-free diaphragm pump. A custom C++ program interfaced with Arduino microcontrollers to control the monochromator and trigger data acquisition. Twenty readings were acquired at each wavelength setting. The nominal 10-layer and 50-layer samples were measured at the University of Texas at Arlington (UTA) using the same VM200 monochromator and deuterium lamp. Transmitted light was detected with a Hamamatsu S8552 VUV photodiode. A Keithley 2450 source meter unit applied a 10 mV reverse bias and measured the resulting photocurrent. The chamber pressure was approximately $10^{-5}$ Torr, and no order-sorting or 130 nm bandpass filter was used. Separate sample, bare-MgF2 reference and dark-current scans were recorded for the 10L and
50L specimens, with 50 repeated readings at each sampled motor position within the reported range. Both single-layer/6-8 layer transmission measurements and 10L/50L transmission measurements end at 325 nm since the Deuterium lamp has negligible intensity in the region from 325 - 400 nm \cite{Saunders1978-os}.

Figure~\ref{fig:transmission} presents all four measured specimen types. For the single-layer and nominal 6--8-layer films, the transmission spectra were digitized from measurement plots shared through internal communication \cite{FebbraroGrapheneDraft} and included for comparison with the 10L and 50L spectra. The shaded bounds for the single-layer and nominal 6--8-layer films, were also extracted from those plots and do not represent statistical uncertainties calculated from the raw measurements. These spectra are normalized to the solvent-processed bare-MgF$_2$ control. The nominal 10L and 50L spectra are calculated from the recorded current scans as described below. Digitized single-layer transmission from Zheng \textit{et al.}~\cite{Zheng2018} is included as a literature comparison in its reported normalization. The digitized data from Zheng \textit{et al.} \cite{Zheng2018} is shown up to 400 nm since the subsequent device measurements were performed near that wavelength. 

\subsection{Normalization, binning and uncertainties}
\label{sec:transmission_analysis}

For the 10L and 50L scans, repeated readings at motor position $j$ were averaged to obtain a current $I_j$ and its standard error of the mean (SEM), $u_j=s_j/\sqrt{n_j}$, where $s_j$ is the sample standard deviation and $n_j=50$ in the reported range. The recorded motor positions $p$ were converted to wavelength using the acquisition calibration,
\begin{equation}
    \lambda=(p-78456)\,(0.017078~\mathrm{nm}).
    \label{eq:wavelength}
\end{equation}
This calibration specifies the wavelength coordinate used in the analysis; an independent uncertainty on its absolute scale is not included in the transmission errors.

Sample, reference and dark currents were binned separately. Within a bin containing $m$ positions, each scan's current and variance of its mean were calculated as
\begin{equation}
    \overline I=\frac{1}{m}\sum_{j=1}^{m} I_j,
    \qquad
    u_{\overline I}^{\,2}=\frac{1}{m^2}\sum_{j=1}^{m}u_j^2.
    \label{eq:binning}
\end{equation}
The bins are 120--125~nm and 2~nm wide from 125 to 325~nm. They include the left boundary and exclude the right boundary. These widths describe analysis bins, not measured monochromator resolution. Transmission is calculated from the binned currents, rather than by averaging position-by-position ratios. Writing the sample, reference and dark current means as $S$, $R$ and $D$, respectively, the reported fraction is
\begin{equation}
    t=\frac{S-D}{R-D},\qquad T=100t~\%.
    \label{eq:transmission}
\end{equation}
This quantity is transmission relative to the bare MgF$_2$ window. It is neither the absolute transmission of freestanding graphene nor that of a completed selenium device.

\begin{figure}[htbp]
    \centering
    \includegraphics[width=\linewidth]{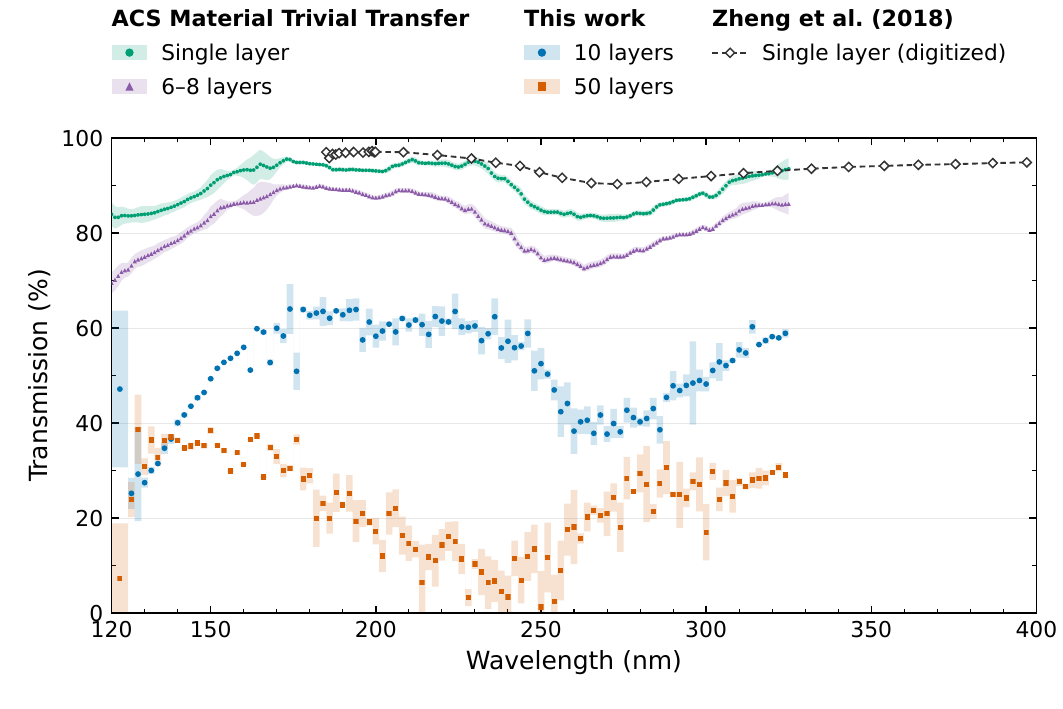}
    \caption{Graphene transmission measurements presented in this work over 120--325~nm, with a single-layer literature comparison that extends to 400 nm. Blue circles and orange squares are the nominal 10L and 50L measurements, respectively, normalized to bare MgF$_2$ using Eq.~\eqref{eq:transmission}; their shaded bands show $\pm1$ propagated SEM and the full bin widths. Green circles and purple triangles show our commercial single-layer and nominal 6--8-layer Trivial Transfer measurements, represented by reconstructed spectra with graphical reconstruction bounds. Open diamonds show digitized single-layer data from Zheng \textit{et al.}~\cite{Zheng2018}, without rescaling; the dashed line is a guide to the eye. All layer counts are nominal. Differences in film preparation, layer count and measurement conditions prevent this comparison from isolating transfer-method performance.}
    \label{fig:transmission}
\end{figure}

The same dark estimate is used in both differences in Eq.~\eqref{eq:transmission}. Assuming independent sample, reference and dark scans and independent position means, first-order propagation gives
\begin{equation}
    u_t^2=\frac{u_S^2+t^2u_R^2+(t-1)^2u_D^2}{(R-D)^2}.
    \label{eq:transmission_error}
\end{equation}
The final term accounts for the covariance introduced by common dark subtraction. The 10L and 50L shaded bands in Fig.~\ref{fig:transmission} show $100u_t$ in percentage points and span the full bin widths. They quantify propagated input SEMs, they do not include calibration errors, source drift, correlations between successive positions, window-to-window differences, or uncertainty from spectral contamination. In particular, they are not a measure of scatter between neighboring wavelength bins. Because no order-sorting filter was used, the spectral purity of these scans is not independently established; the ratios describe the recorded signals at the nominal wavelength settings. The commercial-sample bounds are graphical reconstruction bounds, not SEMs, and no uncertainties are available for the digitized Zheng data.

\subsection{Measured transmission and its implications}

The commercial single-layer and nominal 6--8-layer specimens provide optical access across the measured VUV/UV range: their reconstructed central transmission curves span approximately 83--96\% and 70--90\%, respectively, over 120--325~nm. Both exhibit a broad reduction near 270~nm. These measurements support consideration of the commercial films as transmissive entrance electrodes, while the heating and polymer-removal requirements discussed in Section~\ref{sec:constraint} motivate adapting the transfer procedure for a-Se. The 10L specimen generally transmits more light than the 50L specimen over much of the measured range, both are more attenuating than the commercial single-layer and 6--8-layer specimens over most of their overlapping range. The 10L spectrum shows a broad reduction around 260--280~nm, and the 50L specimen is strongly attenuating through parts of the middle UV range. These observations characterize the transferred optical specimens and motivate further refinement of the transfer procedure, which will be pursued in future work.

Table~\ref{tab:transmission} reports selected measured bins, including the bins near LAr and LXe scintillation. In the 127--129~nm bin, the 10L and 50L estimates are $(29.3\pm9.9)\%$ and $(38.7\pm7.3)\%$, respectively. The two adjacent bins near 175~nm are listed separately to show the local variation around the xenon emission wavelength. These are bin estimates with statistical uncertainties.

\begin{table}[htbp]
    \centering
    \begin{tabular}{lrr}
        \toprule
        Wavelength bin (nm) & Nominal 10L: $T$ (\%) & Nominal 50L: $T$ (\%) \\
        \midrule
        120--125 & $47.2\pm16.5$ & $7.3\pm11.6$ \\
        127--129 & $29.3\pm9.9$ & $38.7\pm7.3$ \\
        173--175 & $64.0\pm5.2$ & $30.5\pm0.7$ \\
        175--177 & $50.9\pm3.9$ & $36.5\pm1.1$ \\
        \bottomrule
    \end{tabular}
    \caption{Selected substrate-normalized transmission estimates. Errors are one propagated SEM, expressed in percentage points; systematic uncertainties are not included. The two rows adjacent to 175~nm are distinct measured bins.}
    \label{tab:transmission}
\end{table}

The 120--125~nm extension requires particular care. It gives $(47.2\pm16.5)\%$ for 10L and $(7.3\pm11.6)\%$ for 50L, the latter is consistent with zero. The dark-subtracted reference signal divided by its propagated SEM is 4.0 and 39.7, respectively. The 10L bin contains five positions in each scan, whereas the 50L sample and reference each contain four positions and the dark scan contains five.  No measurements below 120~nm are shown as the reference becomes poorly resolved in candidate shorter-wavelength bins, making the ratio unstable.

These measurements establish wavelength-dependent optical access through the electrode specimens and quantify the substantial attenuation of the multilayer films. They support further evaluation of graphene entrance electrodes, with thinner films remaining attractive if they can be transferred reliably onto a-Se.

\section{Demonstration of vertical photodetector operation}
\label{sec:photoresponse}

\subsection{Measurement setup}

Both vertical photodetectors were tested in an ARS closed-cycle cryogenic probe station. Light from a 415~nm LED was delivered through a vacuum fiber feedthrough and a 600~$\mu$m-core fiber mounted on a modified probe arm. The illumination spot was contained within the graphene-covered region. A Thorlabs DC2200  \footnote{Mention of specific commercial products or instruments in this 
publication is solely for the purpose of providing a clear description of experimental or test procedures. This 
mention does not constitute an endorsement or recommendation of those specific products by NSLS-II or BNL. It also 
does not imply that the mentioned products are the best available for the intended purpose.} drove the LED with a 50~$\mu$s on time, a 90~ms off time, and a drive current of 100~mA. The optical configuration was kept fixed during the room-temperature and cryogenic measurements.

A battery source supplied positive bias to the Cr bottom electrode. The graphene top electrode was connected to a Tektronix MSO64 oscilloscope with AC coupling and a 1~M$\Omega$ input impedance. Thus, the applied bias and electrical readout used electrodes on opposite sides of the a-Se film. The oscilloscope bandwidth was set to 500~MHz and the sampling rate to $6.25\times10^6$~Samples/s. Acquisition was synchronized to the LED-driver trigger. At each condition, 100 waveforms were averaged by the oscilloscope with only the averaged waveform saved. The displayed traces were baseline-corrected by subtracting the mean voltage between $-1.0$ and $-0.5$~ms relative to the trigger.

The devices were held in the dark for 5~min before acquisition at each condition. During the temperature scan, measurements were taken while warming, with 15~min allowed for equilibration at each set point. Bias values quoted below are nominal source voltages. The recorded observable is the voltage at the AC-coupled readout.

Qualitative blocked-light checks were performed at 293~K and 77~K. The delivery fiber was disconnected at its SMA coupling, covered with opaque metal foil, and reconnected to block light while leaving the LED driver and trigger operating. Device connections, bias, and acquisition settings were unchanged during each check. Visual inspection of the averaged oscilloscope traces showed no discernible trigger-correlated transient above baseline noise when the light was blocked. These qualitative checks support an optical origin for the measured signals.

\subsection{Photoresponse at room temperature}

\begin{figure}[htbp]
    \centering
    \includegraphics[width=0.92\linewidth]{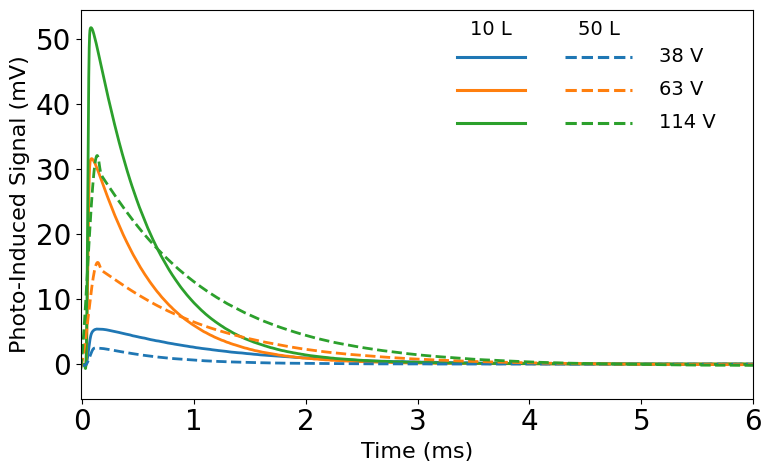}
    \caption{Photoinduced voltage traces at 293~K before the first cooldown. Solid curves show Sample~II (nominal 10L graphene); dashed curves show Sample~I (nominal 50L graphene), at the indicated nominal biases. Each trace is a baseline-subtracted average of 100 optical pulses at 415~nm.}
    \label{fig:room_temperature}
\end{figure}

Figure~\ref{fig:room_temperature} shows the responses at 293~K before the first cooldown, for nominal biases of 38, 63, and 114~V. Both devices produce positive photoinduced voltage transients, and their peak amplitudes increase with applied bias. At 114~V, the displayed peaks are approximately 30~mV for Sample~I (50-L) and 50~mV for Sample~II (10-L).

The opposed-electrode geometry, electrical continuity of the transferred graphene, and bias-dependent signals under illumination establish operation of the completed structures as vertical photodetectors. The devices differ in waveform amplitude and shape, but these differences cannot be assigned uniquely to graphene thickness as the electrode area and transfer history differ, and the recorded voltage includes the response of the readout circuit.

\subsection{Photoresponse at 77 K}

\begin{figure}[htbp]
    \centering
    \includegraphics[width=0.92\linewidth]{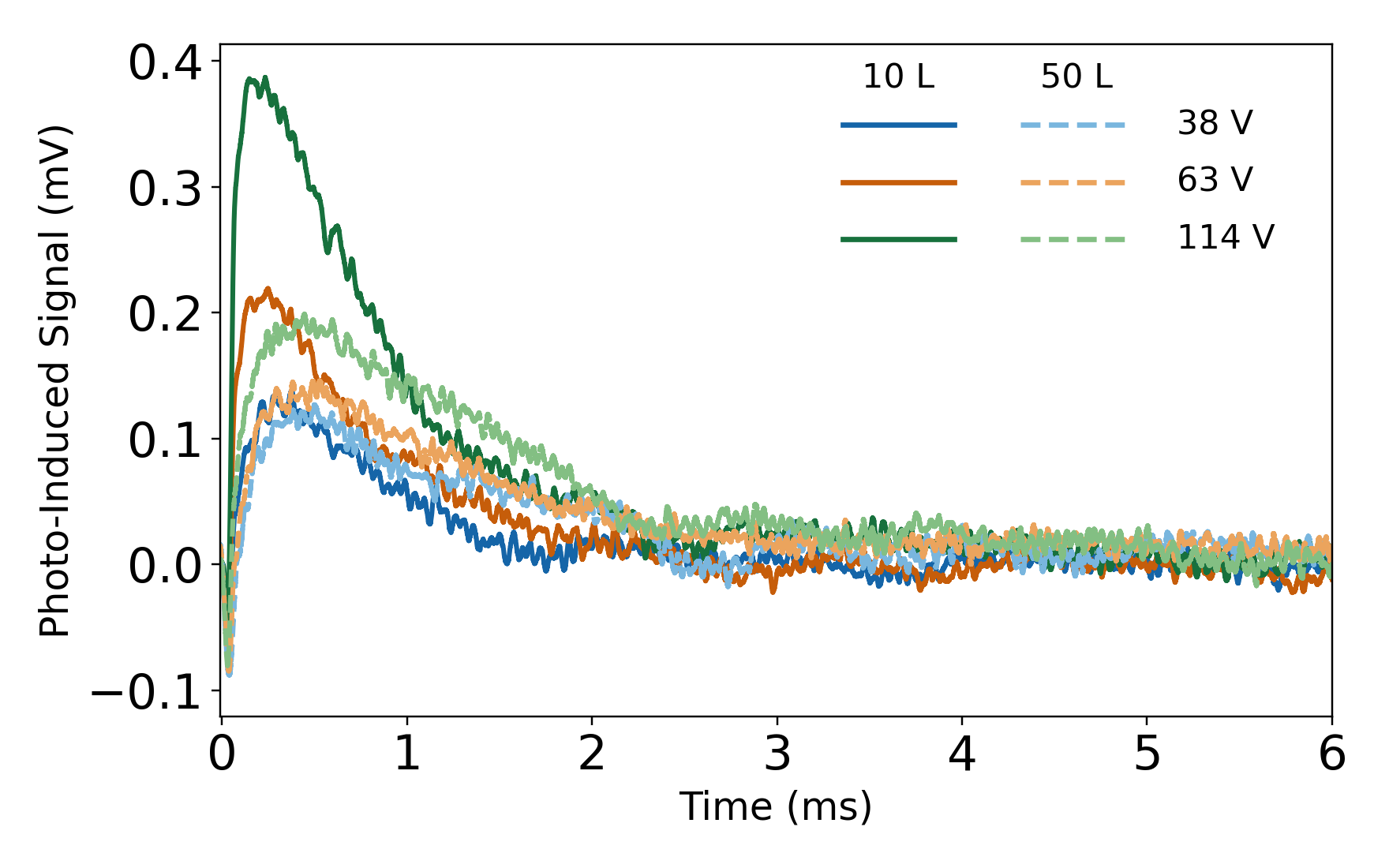}
    \caption{Photoinduced voltage traces at 77~K after the first cooldown and before warming. Solid curves show Sample~II (nominal 10L graphene); dashed curves show Sample~I (nominal 50L graphene), at the indicated nominal biases. Each trace is a saved 100-acquisition oscilloscope average under 415~nm illumination, with the same baseline correction used for the room-temperature data.}
    \label{fig:cryogenic}
\end{figure}

Both devices remained photoresponsive after cooling to 77~K. Figure~\ref{fig:cryogenic} shows the measurements after the first cooldown, before the warming sequence, using the same illumination and readout configuration. Photoinduced signals are visible for both devices, with larger peaks at higher applied bias in the displayed traces. Their amplitudes are smaller than at room temperature. These observations demonstrate that the transferred graphene electrodes support vertical photodetector operation at 77~K.

Measurements during subsequent warming to 293~K also showed photoresponse, although the responses before cooling and after warming were not identical. This demonstrates retained functionality through the tested temperature cycle although it does not establish long-term stability or reproducibility across repeated cycles, which will be tested in future work. 

\section{Discussion and further characterization}
\label{sec:future}

The principal result is a practical change in fabrication of direct graphene transfer for a-Se based vertical devices and vertical photoresponse. The PMMA-free route avoids the sacrificial PMMA support altogether, while the thin-support route reduces its thickness and permits removal using the stated $40\,^{\circ}\mathrm{C}$ acetone treatment. This addresses the thermal and solvent-handling concerns that motivated the work. Raman identification, electrical continuity and photoresponse are the experimental criteria demonstrated here. While additional Raman scans of the completed samples showed no evidence of PMMA or other residual films, a residue-free interface has not been completely established.

For this work, only one device per transfer implementation was studied. The devices differ in nominal graphene thickness, electrode area and processing history, so their signal amplitudes cannot establish a thickness-dependent collection efficiency or rank the transfer procedures. Similarly, the comparison with the commercial Trivial Transfer optical specimens is not a controlled test at fixed graphene thickness. The transmission data provide an empirical optical characterization in place of extrapolations from single-layer spectra.

Future work will focus on the establishment of a fabrication yield using matched devices, quantify structural changes in selenium with low-perturbation measurements before and after transfer, and characterize the graphene/a-Se interface and polymer residue. Calibrated illumination, readout characterization and dark-current/noise measurements are required to determine responsivity, quantum efficiency, detection threshold and intrinsic temporal response. The present AC-coupled voltage waveforms do not independently determine carrier mobility, contact barriers, gain or charge-collection efficiency. Repeated thermal cycling is also needed to assess stability. Finally, direct measurements of the completed vertical devices at 128 and 175~nm are required to test their suitability for noble-liquid scintillation detection.

\section{Conclusion}
\label{sec:conclusion}

We have implemented two graphene-transfer procedures for vertical a-Se photodetectors, motivated by the risk of selenium crystallization during the heating used in conventional Trivial Transfer processing. Nominal 50L graphene was transferred from Ni without a PMMA support, and nominal 10L graphene was transferred from Cu with an approximately 80~nm PMMA support removed in acetone at $40\,^{\circ}\mathrm{C}$. Neither sequence used the commercial $100\,^{\circ}\mathrm{C}$ transfer bake. Raman G and 2D bands identify graphene after processing, and two-terminal measurements establish electrical continuity. The selenium spectra retain an amorphous/disordered contribution, while local structural modification remains unresolved.

We have presented graphene/MgF$_2$ transmission measurements for commercial single-layer and nominal 6--8-layer Trivial Transfer films, together with nominal 10L and 50L films prepared using the adapted routes. The spectra quantify wavelength-dependent attenuation and compare the optical access provided by these electrode specimens. The completed graphene/a-Se/Cr structures produce bias-dependent transients under pulsed 415~nm illumination at 293~K and 77~K. Together, these results establish transfer feasibility and functioning vertical devices. Quantitative detector performance and direct VUV response of the completed structures remain to be measured.

\acknowledgments
This research used the Quantum Material Press (QPress) of the Center for Functional Nanomaterials (CFN), a U.S. Department of Energy Office of Science User Facility at Brookhaven National Laboratory under contract no.~DE-SC0012704. This work is also supported under the Department of Energy Office of Science award DE-SC0022343. The authors also acknowledge and thank the Shimadzu Nanofabrication Facility at the University of Texas at Arlington.

\clearpage
\bibliographystyle{JHEP}
\bibliography{biblio}

\end{document}